\documentclass[10pt,letterpaper]{article}
\usepackage{opex3}
\usepackage{color}
\usepackage{cite}
\usepackage{amsmath}
\usepackage[colorlinks=true,urlcolor=blue,linkcolor=black,citecolor=black,breaklinks=true,bookmarks=false]{hyperref}

\begin{document}

\title{Deep Imaging in Scattering Media with Single Photon Selective Plane Illumination Microscopy (SPIM)}

\author{
Adithya Kumar Pediredla$^{1,*}$, 
Shizheng Zhang$^{1,*}$, 
Ben Avants$^1$, 
Fan Ye$^1$, 
Shin Nagayama$^2$, 
Ziying Chen$^1$, 
Caleb Kemere$^{1,3}$, 
Jacob Robinson$^{1,3,4,\dagger}$, and 
Ashok Veeraraghavan$^{1,5,\ddagger}$
}
\address{
	$^1$ Department of Electrical and Computer Engineering, Rice University, Houston, TX 77005, USA \\
	$^2$ Department of Neurobiology and Anatomy, University of Texas, Medical school at Houston, TX 77030, USA \\
	$^3$ Department of Bioengineering, Rice University, Houston, TX 77005, USA \\
	$^4$ Department of Neuroscience, Baylor College of Medicine, Houston, TX 77030, USA \\
	$^5$ Department of Computer Science, Rice University, Houston, TX 77005, USA \\
	$^*$ Both authors contributed equally to the work
}
\email{$^\dagger$ jtrobinson@rice.edu}
\email{$^\ddagger$vashok@rice.edu} 



\begin{abstract}
	\noindent In most biological tissues, light scattering due to small differences in refractive index limits the depth of optical imaging systems. Two-photon microscopy (2PM), which significantly reduces the scattering of the excitation light, has emerged as the most common method to image deep within scattering biological tissue. This technique, however, requires high-power pulsed lasers that are both expensive and difficult to integrate into compact portable systems. In this paper, using a combination of theoretical and experimental techniques, we show that Selective Plane Illumination Microscopy (SPIM) can image nearly as deep as 2PM without the need for a high-powered pulsed laser. Compared to other single photon imaging techniques like epifluorescence and confocal microscopy, SPIM can image more than twice as deep in scattering media (approximately 10 times the mean scattering length). These results suggest that SPIM has the potential to provide deep imaging in scattering media in situations where 2PM systems would be too large or costly.
\end{abstract}

\ocis{(030.5620) Radiative transfer; (110.0110) Imaging systems; (180.2520) Fluorescence microscopy; (180.6900) Three-dimensional microscopy; (290.0290) Scattering; Selective plane illumination microscopy;} 

\bibliographystyle{osajnl}
\bibliography{SPIM}

\begin{figure}[t]
	\includegraphics[width=\columnwidth]{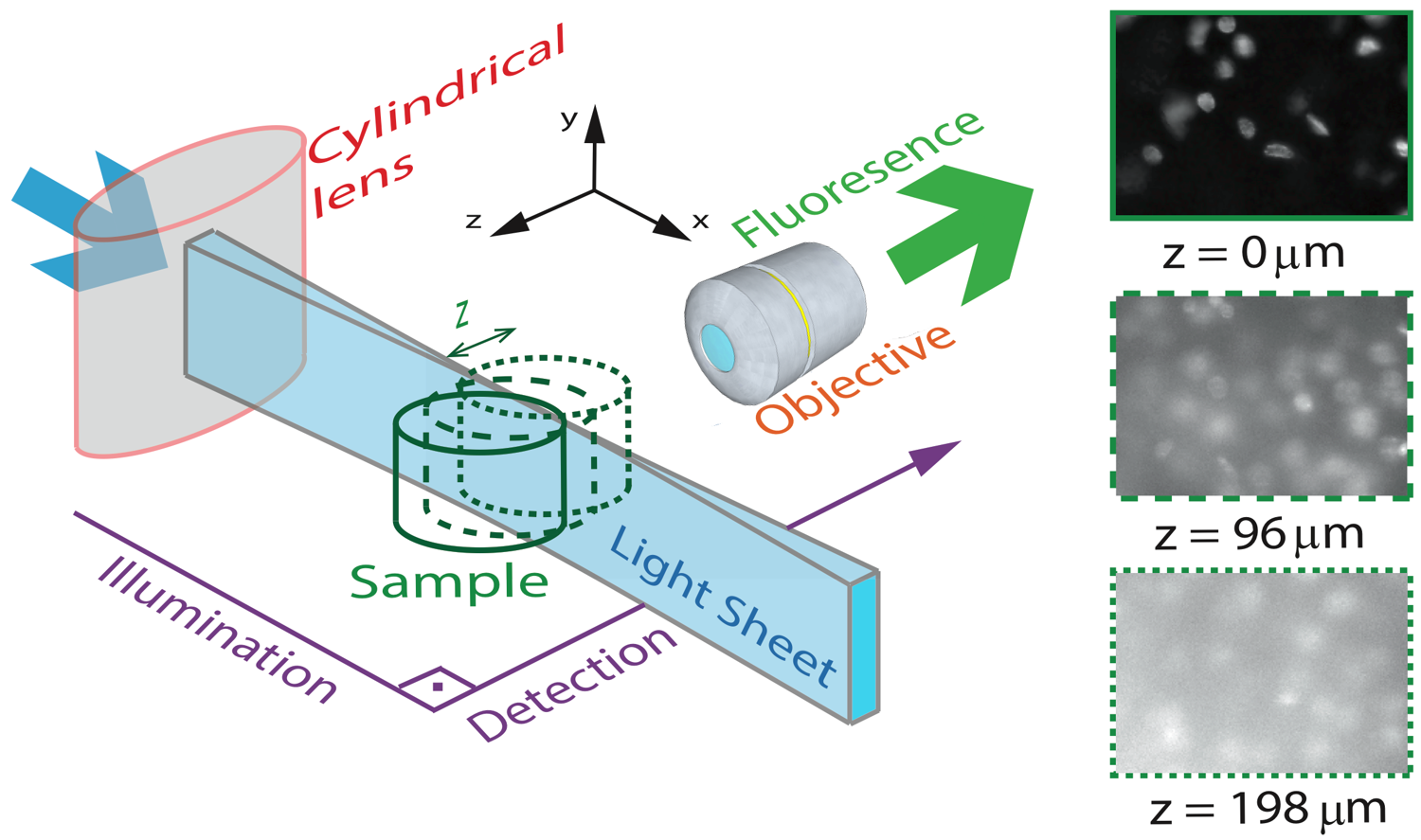}
	\caption{{\bf SPIM Experimental Configuration.} The excitation direction and the fluorescence detection direction are orthogonal to each other. The illumination optics are designed to create a thin light sheet at the focal plane of the imaging objective. By moving the sample in z-direction, one can reconstruct a 3D image of the sample. The right side of the figure shows images at three different depths (z). The quality of image reduces as one images deeper from the sample surface. As a result, there is a limit beyond which the information in the image is completely lost. In this paper, we characterize both theoretically and experimentally this imaging depth limit for SPIM.}
	\label{Fig:SPIMSetup}
\end{figure}
\section{Introduction}
Selective Plane Illumination Microscopy (SPIM) is a 3D fluorescent microscopy technique in which a single planar section of the imaging volume is illuminated and imaged. Scanning plane-by-plane allows one to image the 3D volume of interest. Fig.~\ref{Fig:SPIMSetup} depicts the principle of operation for SPIM. Unlike confocal and multi-photon microscopy, which require point-scanning, SPIM acquires an entire image plane with a single exposure resulting in a significant reduction in the acquisition time required to image an entire 3D volume. The technique was first introduced by Fuchs et al. \cite{fuchs2002thin} in 2002 to image oceanic microbes and has since become a widely used technique for applications that require a fast volumetric acquisition.

SPIM provides two significant advantages over confocal microscopy: (a) faster acquisition times due to plane scanning instead of point scanning and (b) reduced photobleaching of the sample. SPIM is shown to maintain diffraction-limited lateral resolution and high axial resolution (≈ 1.5 $\mu$m) \cite{reynaud2008light}. Due to the advantages mentioned above, SPIM is used extensively for in-vivo microscopy, since the fast acquisition times and reduced photobleaching allow for the imaging of dynamic phenomena. For example, Keller et al. \cite{keller2008reconstruction} and Panier et al. \cite{panier2013fast} recorded nuclei location and movement of zebrafish using a SPIM based multi-view approach. Mertz and Kim \cite{mertz2010scanning} combined the principles of light sheet microscopy and structured light to image whole mouse brain samples. Becker et al. \cite{becker2013ultramicroscopy} imaged macroscopic specimens such as mouse organs, mouse embryos, and Drosophila melanogaster. Jahr et al. \cite{jahr2015hyperspectral} used a diffractive unit to separate the spectral components of the fluorescence (from multiple fluorescence markers) to image Drosophila embryos. More recently, SPIM has been adapted for rapid volumetric imaging in the mouse cortex using a single objective lens \cite{bouchard2015swept}. Despite the growing list of applications, the depth limits of imaging with SPIM are poorly understood. In this paper, we develop a theoretical and experimental basis for characterizing the imaging depth limits of SPIM, and show for the first time that SPIM provides a twofold improvement over confocal and epifluorescence microscopes. We also quantify the maximum imaging depth is affected by light sheet thickness, sensor noise, and fluorophore density. We show that optimally designed SPIM systems can reach imaging depths comparable to 2PM. This deep imaging capability combined with previously demonstrated benefits of fast acquisition and low photo-bleaching make SPIM an attractive technique for biological imaging in scattering tissue.

\section{Related Work}

The depth limits of confocal and multi-photon microscopy is a well-studied property \cite{theer2004fundamental,theer2006fundamental,durr2011maximum,holfeld2007study,wang2011extension,chen2012extending}. Light scattering by tissue is the primary factor that limits the imaging depth: the stronger the scattering, the shallower the depth limit. To meaningfully account for the effect of tissue scattering while comparing the imaging capabilities of various microscopy techniques, depth limits of imaging systems are typically expressed as a multiple of the number of mean free paths (MFPs), or scattering length, within the sample tissue \cite{theer2004fundamental,durr2011maximum}. The mean free path is defined as the average distance a photon travels before a scattering event.

\begin{table}[t]
	\centering
	\begin{tabular}{|c|c|c|}
		\hline
		Microscopy modality & Depth limit & Reference \\
		& (as a multiple of MFP) & \\
		\hline
		\hline
		Confocal & 3-4 & Schmitt et al., 1994 \\
		\hline
		Confocal & 6 & Kempe et al., 1996 \\
		\hline
		2PM & 17 & Theer and Denk, 2006 \\
		\hline
		2PM & 15-22 & Sergeeva et al., 2011 \\
		\hline
	\end{tabular}
	\caption{Comparison of depth limits described in related works. MFP is computed for the emission wavelength, in congruence with \cite{Schmitt:94}, \cite{Kempe:96}.}
	\label{literatureDepthLimits}
\end{table}

This definition of scattering depth limit is independent of various properties of the sample, such as density, the size of the scatterer, and wavelength. Therefore, reporting the depth in terms of MFPs provides a normalized measure to compare depth limits among various microscopic systems and tissue samples.

\subsection{Confocal Microscopy}
Schmitt et al. \cite{Schmitt:94} experimentally demonstrated that the depth limits of confocal microscopy is greater than 3-4 MFPs using a 10-line/mm Ronchi ruling embedded in a suspension of polystyrene latex microspheres in water. Their results show that independent of the ability of the microscope to reject out-of-focus light, there exists a depth limit for confocal microscopy due to the fall of signal intensity. The depth limit is the smallest signal that can be detected by the sensor above the sensor noise floor. Schmitt et al. also showed that decreasing the diameter of the pinhole in the confocal microscope improves the rejection of background scattered light; however, the smaller pinhole comes at the cost of reduced light collection efficiency from the sample plane.

Kempe et al. \cite{Kempe:96} measured the depth limit of confocal microscopy experimentally using a reflecting grating structure suspended in a solution of latex spheres and water. To simulate the effect of imaging deeper, Kempe et al. increased the concentration of the latex spheres thereby increasing the number of mean free paths in their sample. When the concentration of the latex sphere was increased to 2\% (in percent solid content), the contrast of the confocal microscope decreased significantly (around 50\%). Based on these experiments, Kempe et al reported an imaging depth limit of 6 MFPs for confocal microscope. 

\subsection{Multi-photon Microscopy}

In the past decade, two-photon microscopy (2PM) has emerged as the most popular method for deep fluorescence imaging in scattering media due to its superior depth limit compared to confocal microscopy. Theer and Denk \cite{theer2004fundamental,theer2006fundamental} were the first to analytically derive the limits for 2PM as 17 MFP\footnote{MFP length is calculated at emission wavelength of 560 nm. In Theer and Denk’s paper, the result is reported as 7 MFP as they have calculated the MFP length at excitation wavelength.}. They observed that the background fluorescence near the sample surface (i.e. excitation scattering) was the major factor limiting the imaging depth. They noted that as the imaging depth increases, the laser power at the focal point decreases due to scattering of the excitation light, approximately following the Beer-Lambertian law \cite{swinehart1962beer}. To compensate for this loss one must increase the laser power as one images deeper within a scattering media. This increased laser power results in an increase in out-of-focus fluorescence near the sample surface and thus decreases the image contrast. Hence, Theer and Denk defined the imaging depth limit as the depth at which the out-of-focus fluorescence equals the in-focus fluorescence.

As an alternative approach to estimate the depth limit of 2PM, Sergeeva et al. \cite{sergeeva2010scattering} used probabilistic model to estimate the contrast between in-focus and out-of-focus fluorescence. They reported a depth limit between 15-22 MFPs at which the image contrast falls to one \footnote{Sergeeva et al. reported the depth limit as 10-15 MFP, calculated at excitation wavelength, which translates to 15-22 MFPs at emission wavelength of 560 nm.}. 

Both Theer and Denk, and Sergeeva described the depth limit in terms of MFP of the excitation photons − a quantity that is wavelength dependent. Because 2PM and 1PM microscopy use different excitation wavelengths we have chosen to compare the depth limits of 2PM with other imaging modalities in terms of emission MFP (Table~\ref{literatureDepthLimits}). Thus for a given fluorophore the depth limit can be calculated based on the emission wavelength, which is independent of the method of excitation. 

\subsection{SPIM}
Despite the growing popularity of SPIM, the fundamental imaging depth limit of this microscopy technique has yet to characterized either theoretically or experimentally. Paluch et al.~\cite{paluch2013computational} reported that in their experiments the signal strength and the noise strength are equal at 10 MFPs, but did not characterize the factors that affect this maximum imaging depth. Here, using the openSPIM system, \cite{pitrone2013openspim} we also show a depth limit close to 10 MFPs, and take this result further to identify the major factors that influence this imaging depth. 

To computationally estimate the maximum imaging depth for SPIM we employ a Monte Carlo sampling-based approach and compare these results to measured data. We also compare the depth limits of SPIM with confocal, epifluorescence, and two-photon microscopes for brain phantoms and fixed brain tissues. For the first time, our results show that SPIM can image almost twice as deep as confocal microscopy. We also demonstrate that by optimizing the imaging conditions for SPIM, we can approach the imaging depth of 2PM (15 MFPs compared to 17 MFPs). The fact that SPIM has significantly faster acquisition rates compared to point scanning methods, limited photo-bleaching, and deep imaging capabilities, suggests that SPIM should be considered a practical alternative to conventional microscopy techniques in scattering media. 

\section{Problem Statement}
\begin{figure}[t]
	\includegraphics[width=\columnwidth]{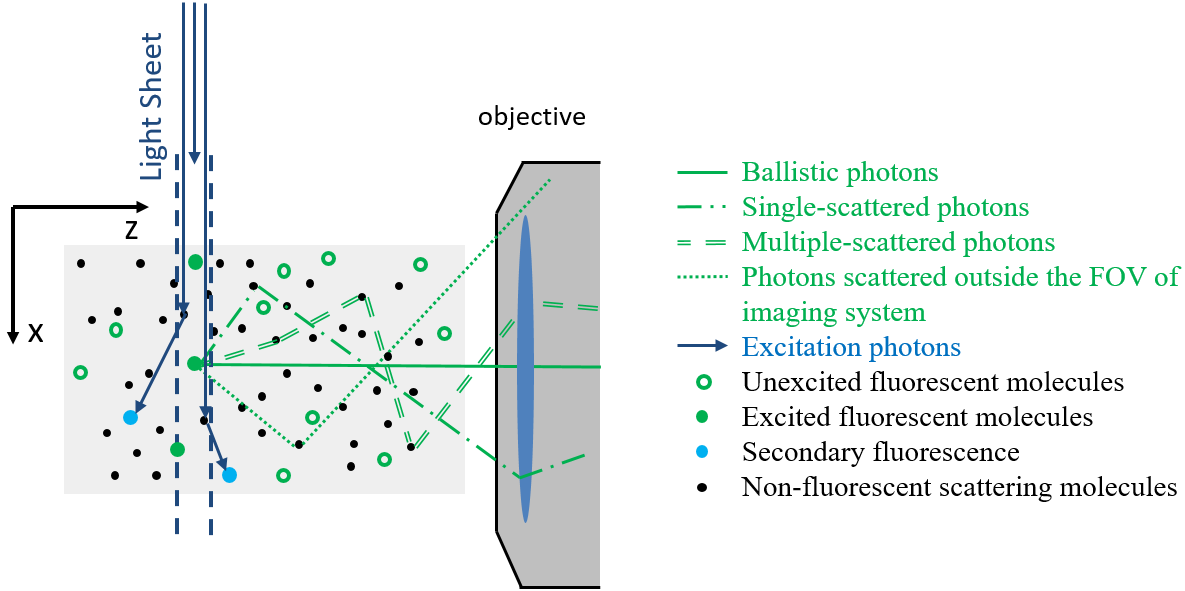}
	\caption{{\bf Monte carlo simulations.} The photons from light sheet that excite the fluorophores are called excitation photons. Some of these excitation photons are scattered and cause secondary fluorescence. 
		The fluorescent molecules that are excited emit photons, referred as emission photons. Some emission photons reach the objective without getting scattered, referred as ballistic photons. Some emission photons reach the objective after single or multiple scattering events, referred as scattered photons. Some emission photons get scattered outside the FOV of the imaging system. 
	}
	\label{Fig:MonteCarlo}
\end{figure}

Scattering of light is the primary factor limiting the imaging depth of most optical microscopes. For fluorescence imaging techniques such as SPIM, scattering can be classified into two types: excitation scattering and emission scattering. {\it Excitation scattering} is defined as the scattering of source photons, while {\it emission scattering} is defined as scattering of fluorescently emitted photons that have a longer wavelength than the source photons. Excitation scattering can also cause {\it secondary fluorescence} as depicted in Fig.~\ref{Fig:MonteCarlo}. Namely, scattered excitation photons can excite unintended fluorophores that then emit light at the emission wavelength. Some of these emission photons, referred as ballistic photons, reach the detector without getting scattered at all, However, many emission photons are either scattered once or more before reaching the imaging system. Because excitation and emission photons are generated by different sources we treat them separately. For SPIM, excitation scattering is independent of the imaging depth because excitation light travels orthogonal to the imaging axis. In other words, at each depth, the excitation light travels the same distance from the source. Excitation scattering results in unwanted background fluorescence, the strength of which can be decreased by decreasing the light sheet thickness. However, the minimum sheet thickness is determined by the diffraction limit of the excitation light. Emission scattering, on the other hand, does depend on the imaging depth. 

Informally, we define the depth limit as the depth below which we cannot distinguish between the signal coming from a specific fluorophore and the signal coming from other sources, such as emission scattered light, secondary fluorescence, and sensor noise. The signal measured by the sensor is influenced by multiple factors including photons arriving from the fluorophore of interest (we will call them photon flux), denoted by ($P$), photons arriving from the background ($B$) due to scattering, Poisson noise of the photon flux and background, quantum efficiency of the sensor ($Q_e$), exposure duration ($t$), sensor read noise ($N_t$), and dark current ($D$). It is easy to robustly estimate the average background signal \cite{cao2007robust}, \cite{chen2006situ}. Hence, we can subtract the background from the measured signal. However, the variance in the background caused by the Poisson noise cannot be removed. Hence, we define signal to noise ratio as 
\begin{align}
	SNR=\frac{PQ_e t}{\sqrt{((P+B)Q_e t+Dt+N_t^2 )}},
\end{align}
\noindent in line with the definition in \cite{Fellers2015}. Formally, we define depth limit as the depth where the signal power and the noise power are equal, alternatively, the depth below which the SNR falls below unity. Note that the SNR is a function of light sheet thickness, fluorophore density, and photon emission rate. In this work, we characterize the depth limit as a function of these parameters – and since many of these parameters have fundamental depth limits themselves, this allows us to estimate the fundamental depth limit for the ideal SPIM microscopes.

\subsection{Assumptions}
We assume a homogenous distribution of scattering particles and fluorophores in the sample. We also assume that the light sheet does not have shadows cast by scattering particles, which could reduce the maximum depth limit.

\section{Computational Model to characterize depth limit}

To characterize the depth limit, we first calculate the fluorophore bead intensity as a function of depth. The depth limit will be the depth where this signal is equal to the total variance created by various sources (Poisson noise, sensor noise, etc.), which we refer to as the visibility threshold. To predict the fluorophore intensity, we employ the classical radiative transfer equation (RTE) \cite{lommel1889photometrie,chwolson1889grundzuge,chandrasekhar1950radiative} given in the standard integro-differential equation for plane-parallel medium as 

\begin{align}
	\cos \theta \frac{dI(\theta, \phi)}{dz} = -\sigma_{ext} I(\theta,\phi)  +  \frac{\sigma_{sca}}{4 \pi}\int_{\theta'=0}^{2\pi} \int_{\phi'=0}^{\pi} \sin \theta' p(\theta', \phi', \theta, \phi) d \theta' d \phi',
	\label{eq:RTE}
\end{align}

\noindent where $\theta$ and $\phi$ are the azimuth and elevation angles, $I$ is the intensity at depth $z$, $\sigma_{ext}$ and $\sigma_{sca}$ are the extinction and scattering cross sections of the scattering particles and $\sigma_{ext}= \sigma_{sca}$ for purely scattering medium, $p(\cdot)$ is the probability (or percentage) of incident radiation traveling in the $(\theta',\phi')$ direction that will shift to $(\theta,\phi)$ direction upon scattering. The first term in the RTE models the amount of light lost due to extinction (absorption + scattering), whereas the second term models the amount of light gained due to scattering in the neighborhood.

While the exact solution to the RTE has yet to be reported, multiple approaches \cite{narasimhan2003shedding,xu2015analytical} to find approximate solutions for the RTE are available in the literature. In this paper, we approximate a solution to Eq.~\ref{eq:RTE} using a Monte Carlo model described below where photon transport from each fluorophore is computed using graphics inspired techniques.

\subsection{Monte Carlo model to solve RTE}
To render the fluorophore, we track many virtual photons emitted by the fluorophore as they pass through the scattering media (see Fig.~\ref{Fig:MonteCarlo}). To track each photon, we model the samples as a random process. For a given scattering particle, photon arrival can be modeled as a Poisson process. We can change our frame of reference such that we model a stationary photon with scattering particles arriving according to a Poisson process. The inter-arrival time for a Poisson process is geometrically distributed \cite{ross2014introduction}. In this case, the arrival time manifests as the distance between two scattering events of the photon. Hence, the distance traveled by the photon before hitting a random scatterer is given by $d = -d_s \log \xi$ , where $d_s$ is the mean free path length and $\xi \sim U[0,\,1]$. After a photon hits a scattering particle, the propagation direction of the photon can change. This deviation is dependent on the scattering medium, and manifests as the $p(\cdot)$ term in the RTE (Eq.~\ref{eq:RTE}). We model this deviation using Henyey-Greenstein probability density function \cite{henyey1941diffuse}, which is given by:

\begin{align}
	\cos \theta = \frac{1+g^2}{2g} - \frac{(1-g^2)^2}{2g(1-g+2g\xi)},
	\label{eq:HGfunction}
\end{align}

\noindent where $g$ is the anisotropic parameter and is around 0.9 for brain tissue \cite{tsia2015understanding}. The azimuth after scattering is uniformly distributed. After scattering, the directional cosines (DCs) of the deflected ray will be altered. We employed the electric field vector, the magnetic field vector, and the direction of the photon as the basis for the 3D space and used the fact that the direction of the electric field before scattering, the direction of the electric field after scattering and the direction of scattered photon are co-planar \cite{namito1993implementation} to find the basis after each scattering event. This approach to track the photon and polarization vectors was first proposed by Peplow \cite{peplow1999direction}. After the photons exit the sample, they are focused to the sensor via lenses in a 4f system. We made the thin lens approximation to model the lens system and track the photon until it reaches the sensor.

The Monte Carlo simulation model that we have proposed in this paper is comparable to the models by Blanca and Saloma for two-photon microscopy \cite{blanca1998monte}, and Schmitt et al. for confocal microscopy \cite{Schmitt:94,schmitt1996efficient}. Our Monte Carlo code, along with the other scripts and data are made available on-line \cite{MonteCarloCode} . Though we consider polarization of the photons, it has to be noted that our Monte Carlo method does not model interference. However, the light emitted by the fluorophore has a typical coherence length of only a few microns \cite{bilenca2007role} and thus the interference effects can be safely neglected at the imaging scales of interest.

\begin{figure}[t]
	\centering
	\includegraphics[width=3.5in]{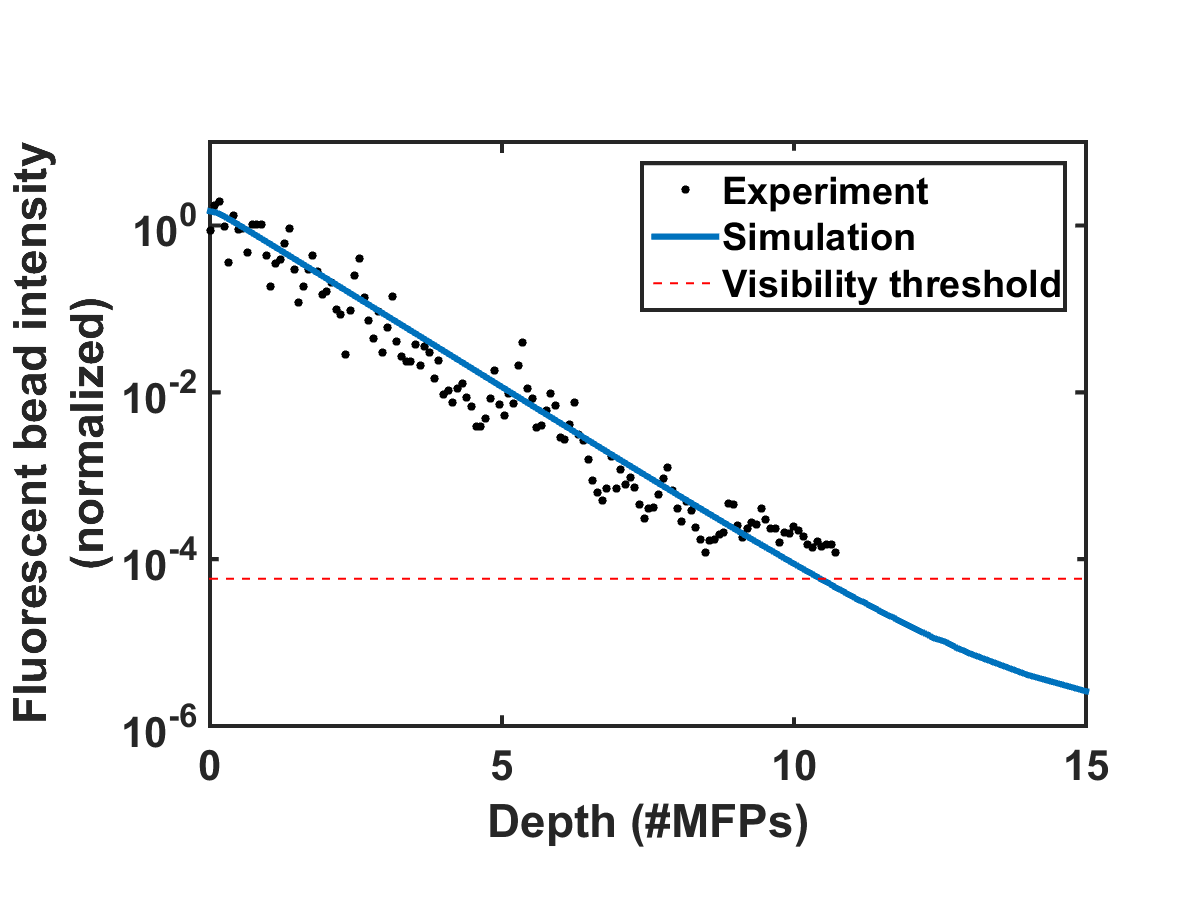}
	\caption{{\bf Monte Carlo model and predicted depth limit.} Simulated (blue line) and experimental (black dot) intensity decay curves for fluorophores imaged with SPIM. We normalize both experimental and Monte Carlo simulated results to have unit intensity at the surface region of the tissue corresponding to an imaging depth of zero to one MFP. Around 9-11 MFPs, the predicted intensity of the fluorophore goes below the visibility threshold (dashed red line) which is equivalent to a SNR of 1. Hence, 9-11 MFPs is the theoretical and experimental depth limit of the open-SPIM used for our analysis.}
	\label{MonteCarlo}
\end{figure}

\subsection{Results and derivation of the theoretical depth limit}

Using the Monte Carlo method, we first estimated the observed intensity (in a SPIM microscope) of fluorophore beads located at various depths in the scattering tissue. The resulting curve is shown in Fig.~\ref{MonteCarlo}. We also experimentally verified the data by imaging fluorescent polystyrene microspheres suspended in a matrix of non-fluorescent microspheres with the approximate scattering properties of brain tissue (i.e. brain phantom, see methods). Since the intensity of fluorophores we measure is a function of exposure duration, we cannot directly compare the experimental fluorophore intensities with the simulation results. Hence, we normalize both experimental and Monte Carlo simulation results to have unit intensity at the sample surface, which corresponds to an imaging depth of less than one MFP. From Fig.~\ref{MonteCarlo}, we observe that the experimental fluorophore bead intensities at various imaging depths are close to the intensities predicted using our model. We have experimentally measured the visibility threshold near the depth limit for a Hamamatsu sensor (type number: C4742-95-12ER). Recall that the theoretical depth limit is the depth where the Monte Carlo predicted bead intensities fall below this visibility threshold. Using the values from the Monte Carlo simulation (Fig.~\ref{MonteCarlo}), we can conclude that the theoretical depth limit is between 9 and 11 MFPs.

It should be noted that this particular method for finding the depth limit is informed by experimental results that set the visibility threshold. The background variance part of the visibility threshold is dependent on vortex mixing during the sample preparation, preventing us from calculating the visibility threshold. However, once we know the visibility threshold for a particular SPIM system, we can predict this threshold for other SPIM systems by scaling the threshold appropriately (according to physical parameters of the system such as light sheet thickness, fluorophore density, sensor characteristics, etc.). In the next subsection, we investigate how the light sheet thickness and sensor properties affect the depth limit for SPIM. 

\begin{figure}[t]
	\centering
	$
	\begin{array}{c}
	\includegraphics[width=3in]{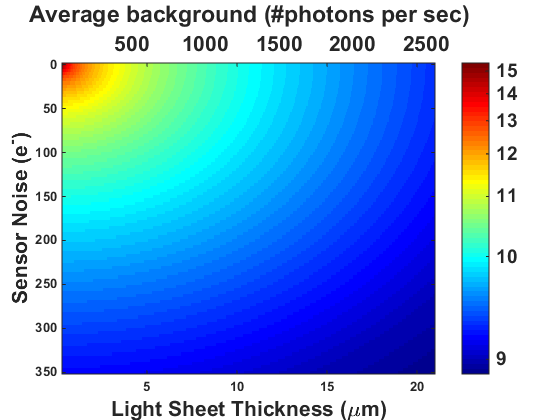} \\
	\mbox{(a)}
	\end{array}
	$
	$
	\begin{array}{cc}
	\includegraphics[width=2.5in]{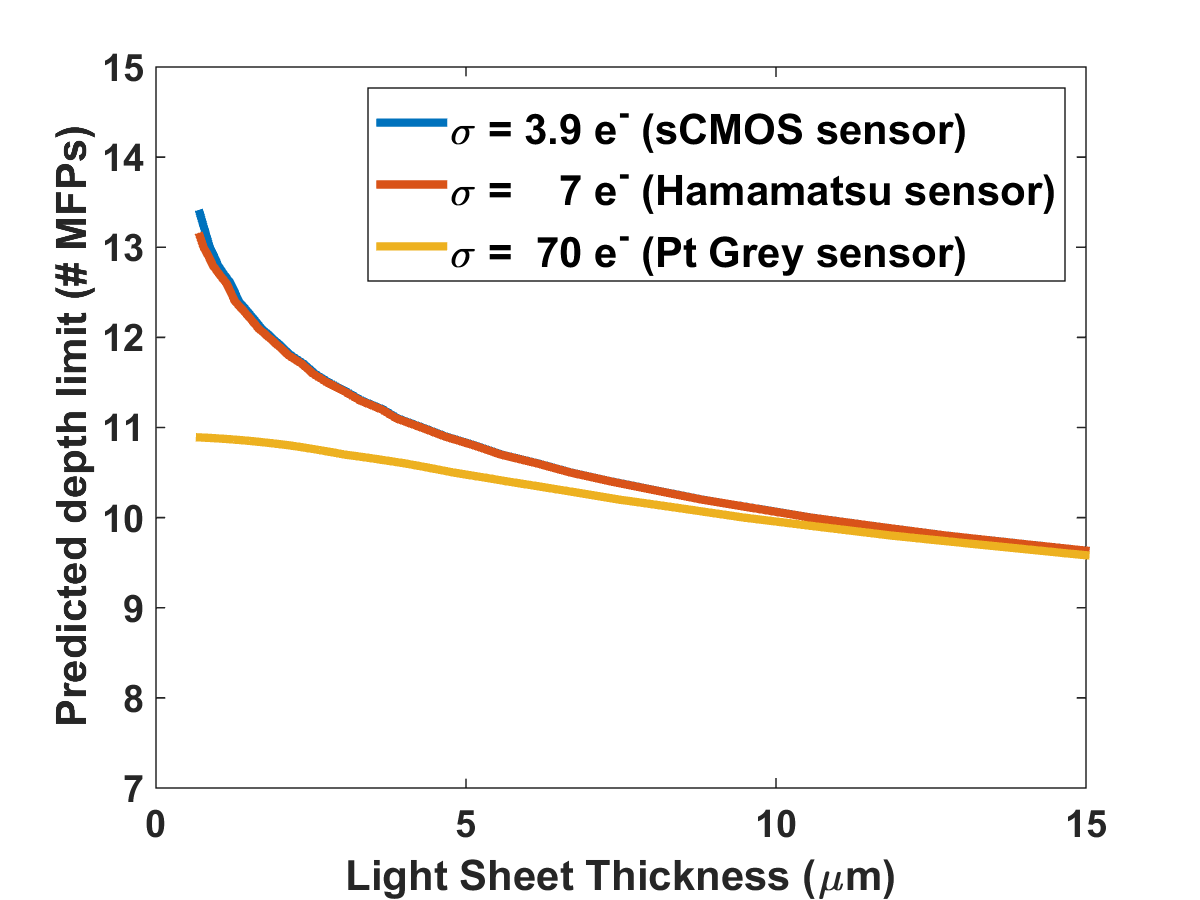} &
	\includegraphics[width=2.5in]{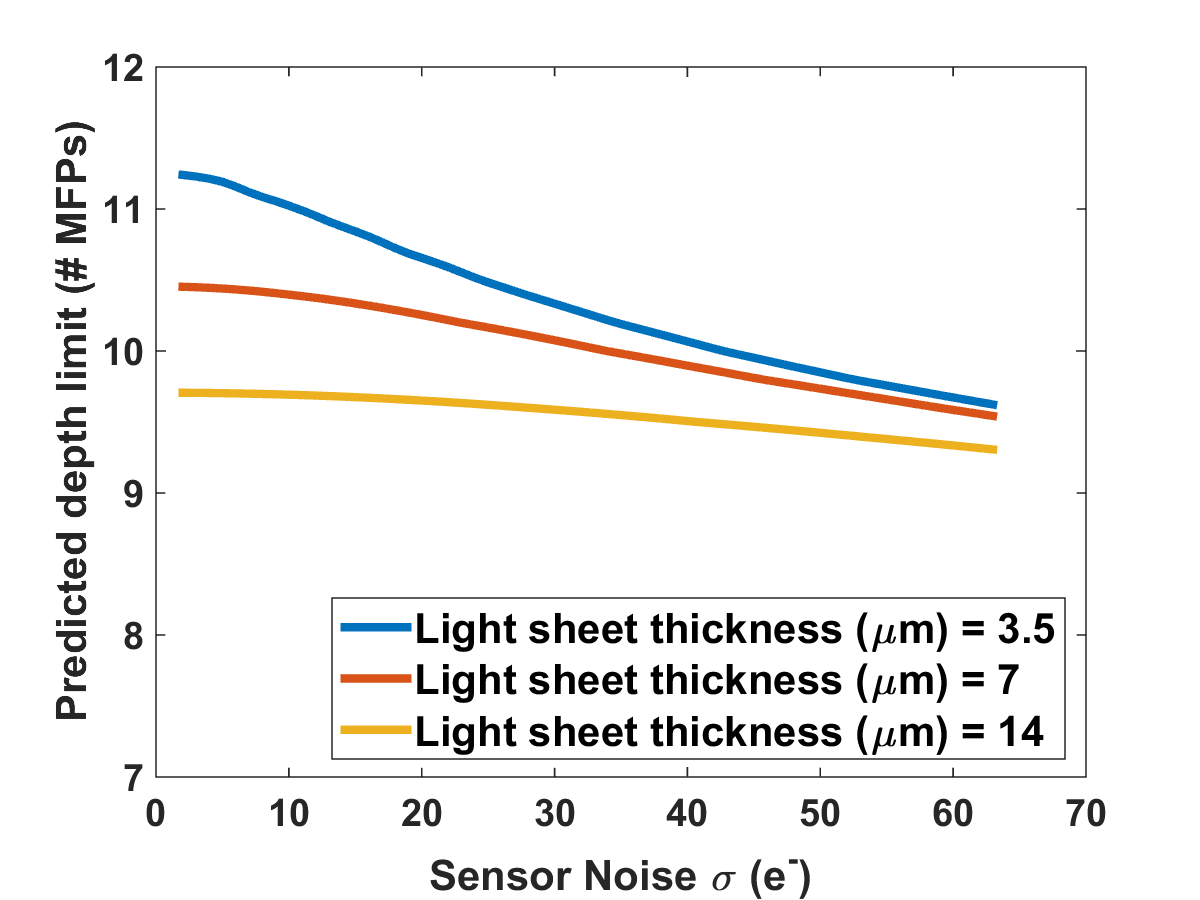} \\
	\mbox{(b)} & \mbox{(c)} 
	\end{array}
	$	
	\caption{{\bf Characterization of depth limit.} (a) Depth limit as a function of light sheet thickness and sensor noise. The axis label “Average background” represents the average number of background photons per second for our SPIM system. This value is obtained by multiplying the light sheet thickness with the field-of-view of a pixel (1$\mu$m $\times$ 1$\mu$m), fluorophore concentration (61 million molecules per ml), and fluorophore emission rate (2 million photons per molecule per second \cite{zhao2011photon}). (b) Predicted depth limit for different commercially available sensors for various light sheet thicknesses. Independent of the sensor, as the light sheet thickness decreases, we observe an increase in the depth limit. (c) Predicted depth limit for various sensor noise levels at a given light sheet thickness. We observe a decrease in the depth limit as the sensor noise increases.}
	\label{Fig:DepthLimitCharacterization}
\end{figure}

\subsection{Depth limit characterization}
The depth limits of a particular SPIM system is primarily determined by the light-sheet thickness and sensor noise level. Therefore, we characterize the depth limit as a function of light sheet thickness, and sensor noise. As the light sheet thickness increases, the amount of background fluorophore light increases linearly. Therefore, an increase in light sheet thickness results in an increase in the variance of the background and thereby contributes to an increase in the visibility threshold. The increase in visibility threshold decreases the depth limit of imaging according to the graph shown in Fig.~\ref{MonteCarlo}. Increasing the fluorophore density has a similar effect as increasing the light sheet thickness.

When we use a sensor with lower noise characteristics, it is natural to expect that the imaging depth limit would increase. Formally, this effect appears in Eq.~\ref{eq:RTE} as a decrease in the visibility threshold as the sensor noise is reduced. Similarly, an increase in fluorophore emission rate is also expected to increase the imaging depth limit as each fluorophore will emit more photons, increasing the signal intensity. Assuming that the fluorophore emission rate has increased by a factor of $\kappa$ ($\kappa > 1$), both the photon flux and the background flux are increased by the same factor of $\kappa$. The new SNR increases approximately by a factor of $\sqrt{\kappa}$ and is given by 

\begin{align}
	SNR=\sqrt{\kappa }\frac{PQ_e t}{\sqrt{(P+B)Q_e t+\frac{Dt+N_t^2}{\kappa} }}.
\end{align}

\noindent The increase in SNR will result in a corresponding increase in the depth limit of the SPIM. 

To calculate the imaging depth limit for a general SPIM system, we can scale the background variance according to the light sheet thickness, and we can adjust the sensor noise characteristics. By rescaling Fig.~\ref{MonteCarlo} in this way, we compute the depth limit for arbitrary sheet thicknesses and sensor performance. We plot this result in Fig.~\ref{Fig:DepthLimitCharacterization}, which shows the expected depth limit for various light sheet thicknesses and sensor noise levels. We notice that as the light sheet thickness or the sensor noise levels increases, the depth limit decreases. It should also be noted that this depth limit graph is only valid if other parameters like the field-of-view, fluorophore density, and fluorophore emission rate are held constant. To make this fact more explicit we rewrite the light sheet thickness more generally as the number of background photons per second and add this axis to the top of Fig.~\ref{Fig:DepthLimitCharacterization}. The background photon rate is calculated as the product of the field-of-view of a pixel in the sensor (1$\mu$m $\times$ $1\mu$m), light sheet thickness, fluorophore concentration (61 million molecules per ml), and fluorophore emission rate (2 million photons per molecule per second \cite{zhao2011photon}).

\begin{figure}[t]
	\centering
	\includegraphics[width=3.5in]{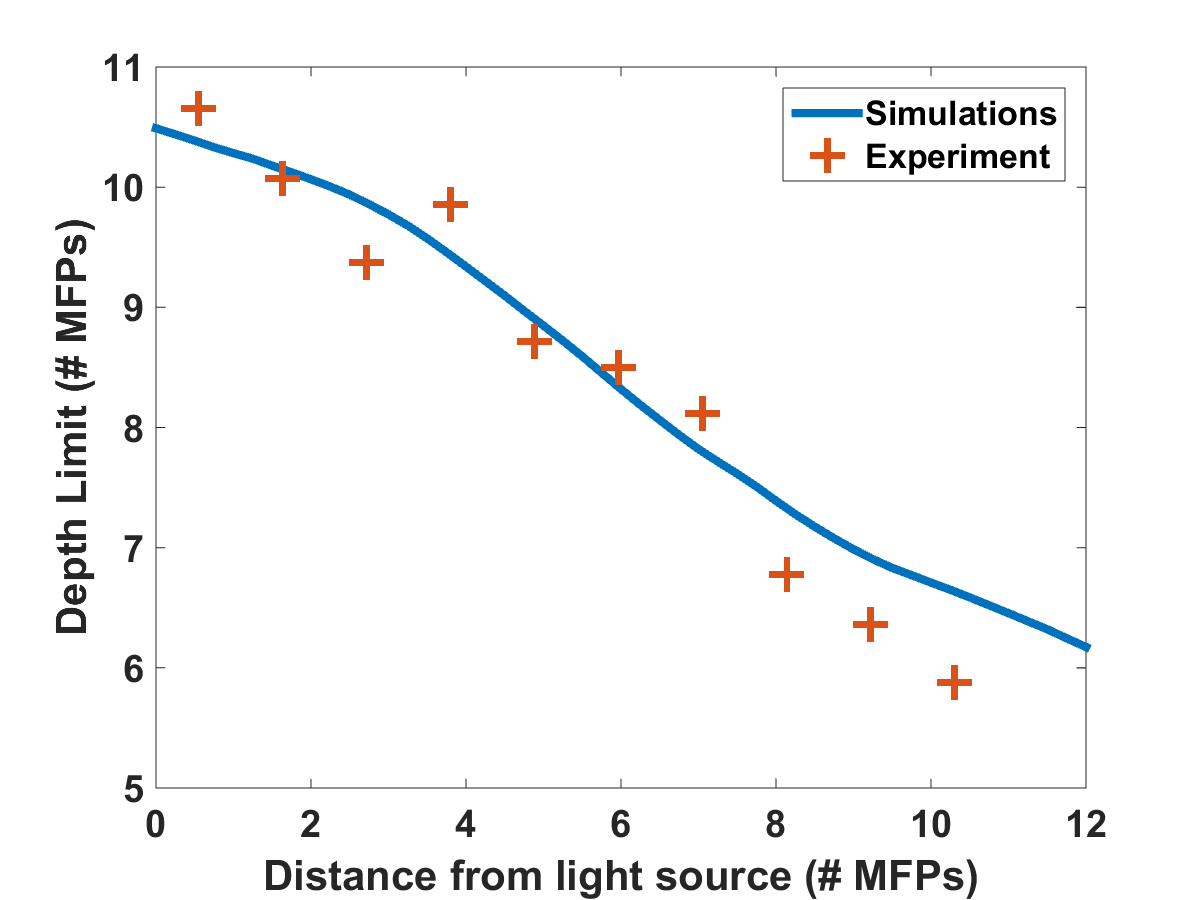}
	\caption{{\bf Validation of depth limit characterization.} We compute the light sheet thickness as a function of distance (z-direction) from the light source using Monte Carlo simulations. Using the depth-limit characterization graphs, we compute the corresponding depth limit as a function of distance from the source, which is represented by the blue curve. We have also measured the depth limit experimentally and plot this limit as red crosses. Notice that the depth limits estimated by our technique and the observed depth limits match with a maximum error of 1 MFP.}
	\label{Fig:DepthLimitValidation}
\end{figure}

\subsection{Validation of Depth limit characterization:}
To verify the predicted dependence of the depth limit on light sheet thickness we analyzed the imaging depth as a function of the distance from illumination source. We chose this experiment because the incoming light-sheet illumination expands as the light propagates through sample. Thus the light sheet thickness expands uniformly as a distance from the source giving us the ability to test our model using the natural expansion of the light sheet without physically altering the experimental setup. Moreover, this kind of validation also enables us to understand the effect of imaging farther from the excitation source (x-direction in Fig.~\ref{Fig:SPIMSetup}).

To estimate the light sheet thickness at a distance ‘x’ from the illumination source, we used Monte Carlo simulations. The photons are initialized on a thin light sheet and are uniformly distributed along y-axis direction and Gaussian distributed along z-axis direction with full-width-half-maxima (FWHM) equal to the light sheet thickness of our SPIM system ($\sim$7 $\mu$m). To generate the initial z-coordinate for photons, we used a Box-Muller transformation \cite{box1958note}. The sensor (virtual) is placed parallel to yz-plane at a distance ‘x’. After computing the light sheet thickness, we queried the depth limit characterization plot to estimate the depth limit as a function of the distance from the light source and plotted Fig.~\ref{Fig:DepthLimitValidation}. The blue curve shows the estimated depth limit as a function of the distance from the light source. We plotted the empirically measured values in red. We observe that the estimated depth limits follow our experimental results with a maximum deviation of 1 MFP confirming our calculations for how the maximum SPIM imaging depth depends on light sheet thickness.

\section{Comparisons of depth limits with Epifluorescence, Confocal, and 2PM microscopes}
Epifluorescence, Confocal, and 2PM microscopes are the most common volumetric microscopy techniques. In this section, we compare the depth limits of SPIM with these common imaging modalities. To perform this comparison, we prepared a series of brain phantoms and fixed brain tissue, and imaged these samples using the aforementioned four microscopes: an epi-fluorescence microscope, a confocal microscope, a SPIM microscope, and a two-photon microscope. The section ``Methods" describes the preparation of samples and the procedures used to image them. In this section, we compare the results. 

\begin{figure}[t]
	\centering
	\includegraphics[width=\columnwidth]{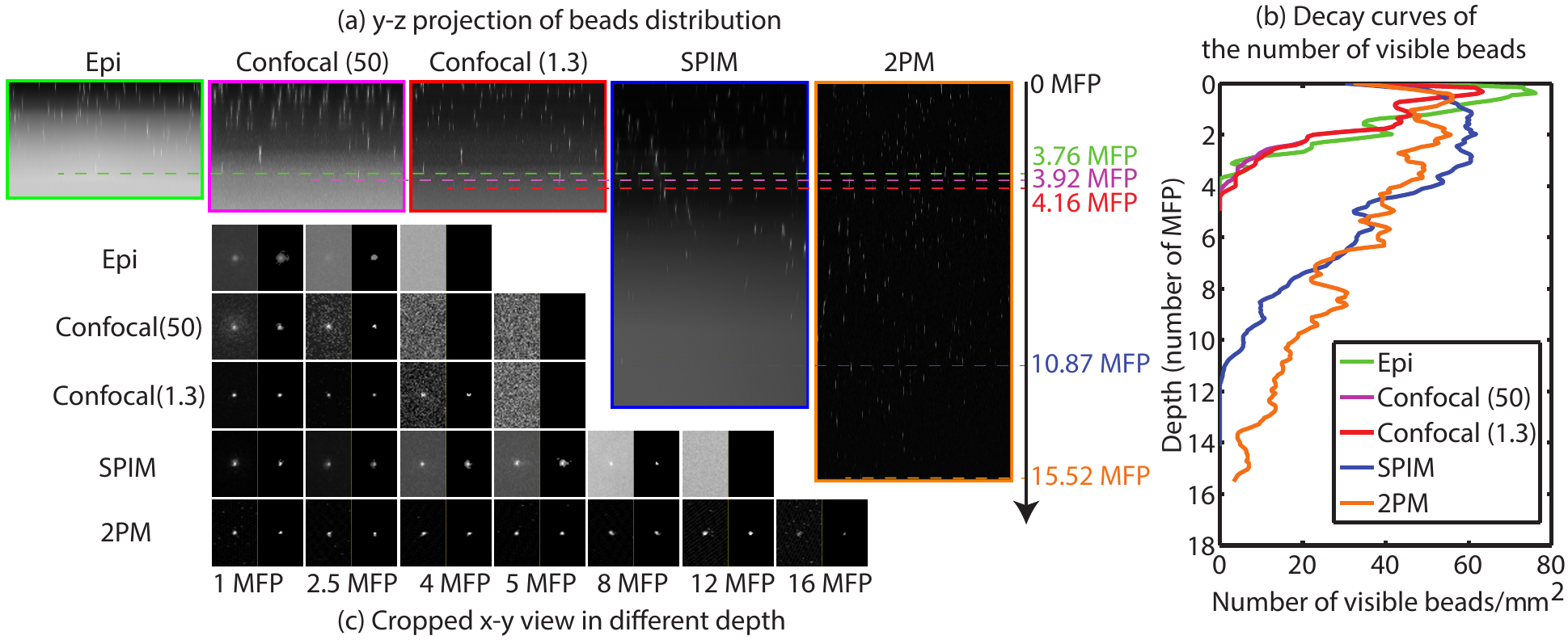}
	\caption{{\bf SPIM vs other microscopies for brain phantom.} (a) y-z maximum intensity projection of the 3d-volume of brain phantom as imaged by epifluorescence microscopy, confocal microscopy ($\nu$ = 50 optical units), confocal microscopy ($\nu$ = 1.3 optical units), SPIM, and two-photon microscope. (b) The number of visible beads per square millimeter at each depth for the five different microscopy methods. (c) cropped x-y images of individual beads at various imaging depths from each of the microscopy techniques. Images are zoomed-in (4X) for each bead. All x-y slice images show both post-processed images with adaptive histogram equalization to improve the visual quality (left), and raw data (right).}
	\label{Fig:BrainPhantom}
\end{figure}
\begin{table}[t]
	\centering
	\begin{tabular}{|c|c|c|c|}
		\hline
		Density compared to  & MFP & Depth limit ($\mu$m) & Depth limit average \\
		brain phantom &&&( \#MFPs) \\
		\hline
		\hline
		100\% & 75.03  & 807.89  & 10.64 \\
		\hline
		50\% & 140.12 & 1422.21 & 10.15 \\
		\hline
		25\% & 238.53 & 2199.24 & 9.33 \\
		\hline 
	\end{tabular}
	\caption{Experimental depth limits of SPIM for various brain phantoms.}
\end{table}
\begin{table}[t]
	\centering
	\begin{tabular}{|c|c|c|}
		\hline
		Imaging modality & Depth limit ($\mu$m) & Depth limit average \\
		&&( \#MFPs) \\
		\hline
		\hline
		Epifluorescence & 255.10  & 3.40 \\
		\hline
		Confocal ($\nu$=50) & 297.87 & 3.97 \\
		\hline
		Confocal ($\nu$=1.3) & 335.38 & 4.47 \\
		\hline
		SPIM & 799.32 & 10.64 \\
		\hline
		2PM & $>$1164 & $>$15.51 \\
		\hline
	\end{tabular}
	\caption{Depth limit comparison of various imaging modalities for brain phantom.}
\end{table}

\subsection{Comparison of depth limits of various microscopy techniques on brain phantoms}

To compare 3D microscopy techniques we prepared a tissue phantom with the approximate scattering properties of the brain using a suspension of fluorescent and non-fluorescent polystyrene microspheres (see Methods). Fig.~\ref{Fig:BrainPhantom} (a) shows x-y images of individual beads captured at different depths using each of the microscopy techniques. Notice that the depth limits of epi-fluorescence and confocal microscopes are less than 5 MFPs, whereas SPIM can image between 9 and 11 MFPs. We repeated the experiment on multiple samples and with brain phantoms of different densities. Fig.~\ref{Fig:BrainPhantom} summarizes the depth limits of various brain phantoms. We can observe that the ratio of the depth limit to MFP length, which we expect be a constant, is slightly lower for larger MFP length. This decrease in the depth limit for large MFPs has also been observed in two-photon microscopy \cite{durr2011maximum}. For long MFPs the total distance traveled by the emitted photons is greater, thus we expect that other loss mechanisms like absorption may contribute to lower shallower than expected depth limits.

Fig.~\ref{Fig:BrainPhantom} also shows a y$-$z projection of the images collected from the brain phantom using an epifluorescence microscope, a confocal microscope (pinhole sizes of 1.3  and 50 optical units), a SPIM and a 2PM. We also counted the number of visible beads per image frame and plotted this number as a function of imaging depth (averaged over four samples). We observe that, at the surface region of the phantom, the bead counts of all imaging modalities are almost the same. However, the bead count quickly drops as we image deeper using epifluorescence and confocal microscopy. Epifluorescence imaging appears to have slightly more visible beads because this method can observe more out-of-focus beads than any of the other imaging techniques. Hence, even though epifluorescence imaging shows more beads, they are not necessarily located at the depth of the focal plane. As expected, confocal microscopy with an optimized pinhole size (1.3 optical units) has better depth resolution than confocal microscopy with a large pinhole size (50 optical units). Thus, the depth limit of confocal microscopy with the optimized pinhole size is higher than the confocal microscopy with large pinhole size. However, both methods show depth limits between 3$-−$4 MFPs. SPIM has a depth limit of 9$-−$11 MFPs, which is 2$-−$3 times deeper than other single-photon-microscopy modalities. 2PM can image even deeper than SPIM, most likely due to the fact that 2PM can better confine the excitation volume and thus reduce the background fluorescence.  Due to the limited working distance of our 2PM, we can explore a maximum depth of around 1160 $\mu$m (15.52 MFPs) before the objective lens contacts the sample surface. At this maximum depth $-$ determined by the working distance of the objective lens $-$ the beads have very low intensity. Thus, we predict that the imaging depth limit is slightly longer than 15.52 MFPs.

The factors that limit the maximum imaging depth is different for each imaging modality. Epifluorescence imaging suffers from both excitation scattering and emission scattering, and thus the images appear less sharp with bright background levels. Confocal imaging blocks most emission scattering, but still suffers from excitation scattering. Additionally, near the depth limit, the confocal system has very little light that reaches the sensor through the pin hole. Thus the count noise on the image sensor becomes a significant factor that limits the imaging depth. SPIM effectively eliminates most of the effects of excitation scattering. Because this method suffers primarily from emission scattering alone, it has a much deeper maximum imaging depth compared to other single-photon-excitation microscopy techniques. In these samples, the depth limit of SPIM is likely limited by the finite sheet thickness, which does not confine the excitation volume as well as 2PM. We expect that in the limit of a diffraction limited light sheet, 2PM and SPIM would have the same depth limit ($\sim$15 MFPs) as shown Fig.~\ref{Fig:DepthLimitCharacterization}(a).

\begin{figure}[t]
	\centering
	\includegraphics[width=\columnwidth]{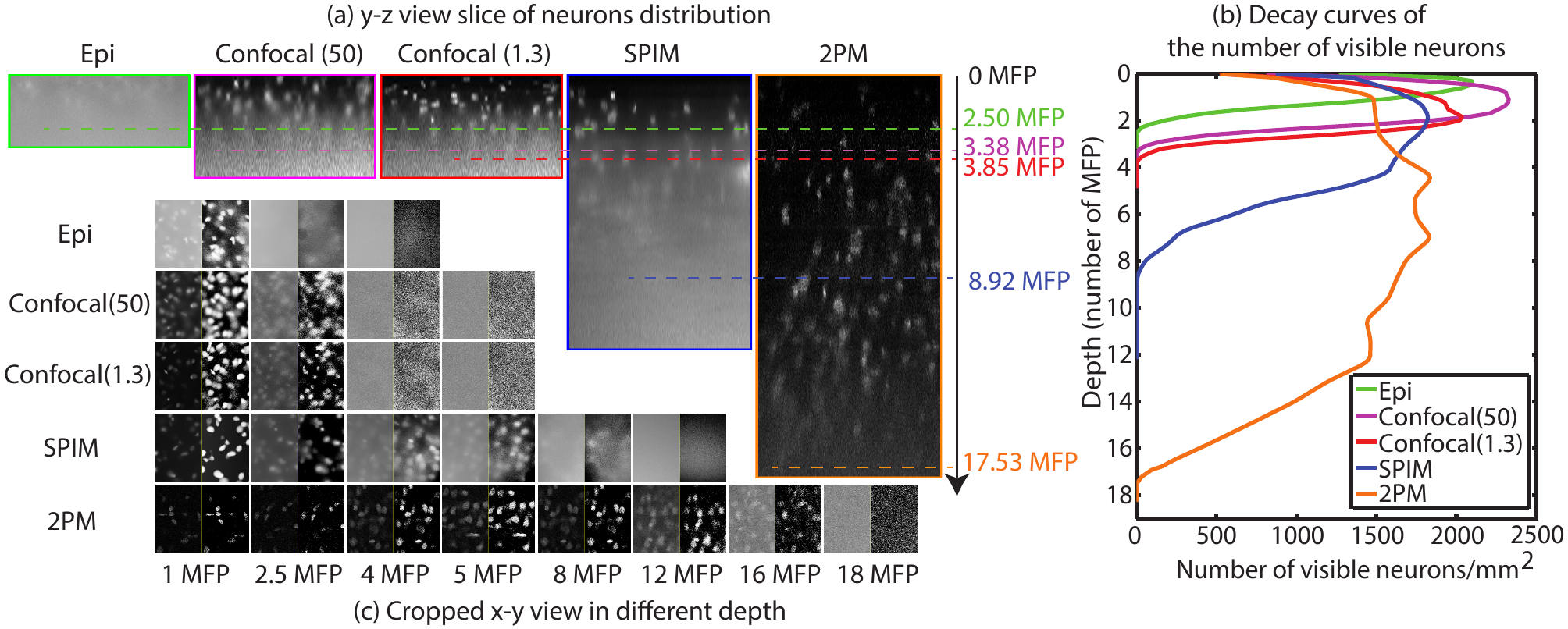}
	\caption{{\bf SPIM vs other microscopies for fixed brain tissue.} (a) Figure is y - z view slices of the fluorescent bead distribution in a sub-volume of brain tissue by epifluorescence microscopy, confocal microscopy ($\nu$ = 50 optical units), confocal microscopy ($\nu$ = 1.3 optical units), SPIM and two photon microscope. (b) Plot shows the corresponding decay curves of the number of visible neurons per 1 square millimeter on camera frame. (c) Figure shows cropped x-y view images of camera frame at various imaging depths in different modalities. All x-y slice images are post-processed with adaptive histogram equalization to improve the visual quality, as shown on the right side of the raw images.}
	\label{Fig:BrainTissue}
\end{figure}
\subsection{Comparison of depth limits of various microscopy techniques on brain tissue}
We also compared the depth limits of the major 3D fluorescent microscopy techniques using a fixed brain tissue sample isolated from the cerebral cortex of mature male Long-Evans rat. After performing perfusion and fixation by PFA (4\% paraformaldehyde in buffer), the sample becomes denser than live brain tissue. Then it is stained in diluted NeuroTrace 500/525 green fluorescent Nissl stain. Note that the fixation process greatly increases the amount of light scattering in the tissue sample leading to a MFP that is reduced by roughly a factor of two compared to live tissue. Fig.~\ref{Fig:BrainTissue} shows both the x$-$y view at different depths and the y$-$z view of the cerebral cortex as imaged by epifluorescence microscope, confocal microscope ($\nu$ = 50 optical units), confocal microscopy ($\nu$ = 1.3 optical units), SPIM, and a two-photon microscope. In an x$-$y slice, we can notice neurons as bright ellipses. We post-processed the x$-$y slices with adaptive histogram equalization method to improve the visual quality of the slices. The behavior of signal intensity with respect to background as we image deeper is similar to that of brain phantom. We also counted the number of visible neurons per image frame and plotted the decay curve of visible neurons per image as a function of imaging depth. One difference from brain phantom's decay curve is that the decay curve is relatively flat at mid depths and then drops suddenly at large depths. This phenomenon is likely due to the fact that a 1 $\mu$m bead in the brain phantom only occupies few pixels and is difficult to identify as it blurs. However, for brain tissue, a 10 $\mu$m neuron occupies a many pixels, and can therefore be identified upon blurring. This phenomenon also accounts for a smoother variation in distinguishable neurons compared to the number of beads visible in the brain phantom (Fig.~\ref{Fig:BrainPhantom}). Based on these measurements we identified the depth limits of the epifluorescence, confocal, SPIM, and two-photon microscopes for fixed brain tissue sample as 2.50, 3.38, 3.85, 8.92, and 17.53 MFPs respectively, which are similar to results of the experiments in the brain phantom.

\section{Methods}
Here we describe the experimental methods for preparing brain phantom and brain tissue as well as the procedure used to measure the mean free path (MFP). We will also give details of various steps involved in imaging these samples with confocal, epi-fluorescence, two-photon, and SPIM microscopes. 

\subsection{Preparation of samples}
In this section, we give comprehensive details employed for the preparation of various samples that can help researchers in replicating the samples for imaging.

\subsubsection{Preparation of brain phantom}
To prepare brain phantom for imaging, we suspend blue-green fluorescent (1 $\mu$m diameter, 488/560) and non-fluorescent polystyrene beads (1 $\mu$m diameter) in agar/milli-Q solution. Low melting agar powder is measured out to be 1\% wt/v of the final sample volume and added to the milli-Q water. The milli-Q water is heated to boiling to activate the agar. Then the milli-Q/agar solution is vortexed (spun on a vortex spinner) to make agar evenly mixed up. Next we add fluorescent and non-fluorescent polystyrene beads into the milli-Q/agar solution. Fluorescent and non-fluorescent beads are stored separately in a suspension which has   beads per mL. The proportion of milli-Q water to sample suspension is calculated based on the initial bead suspension concentration and the target concentration of the sample. The beads cannot be heated above 97 degrees Celsius, so the milli-Q/agar solution is cooled down to 60 $-$ 85 degrees Celsius before the beads are added. After adding the beads, the sample is vortexed to ensure that the beads are evenly distributed in sample suspension before it is transferred into syringes and cuvettes. These containers are refrigerated for several minutes to make the sample coagulated. We made sure that the samples across various experiments like measuring mean free path, computing experimental depth limits of SPIM, and other imaging modalities come from the same batch. For the brain phantom, in the final sample volume, we have about    non-fluorescent beads per mL as scatterers to simulate brain tissue and about   fluorescent beads per mL as features to measure depth limit.

\subsubsection{Preparation of brain tissue}
To prepare brain tissue for imaging, we fixed and extracted the brain\footnote{The sample was obtained from animals euthanized	as a result of other procedures approved by the Institutional Animal Care and Use Committee at Rice University conforming to National	Institutes of Health guidelines.} of an adult Long Evans rat and stained neurons with a green fluorescent dye. The animal was transcardially perfused with 4\% paraformaldehyde in phosphate-buffered saline (PBS) solution. After further fixation, the brain was manually divided into approximately 1 mm coronal sections using a brain slicer matrix. Sections were incubated for 24 hours at 4$^{\circ}$ C in a solution containing a NeuroTrace 500/525 fluorescent Nissl stain (Molecular Probes, 1:200 dilutions in 0.1M PBS). The stained sections were then cut into approximately 1 mm strips along the horizontal plane and divided at the midline, with samples from either hemisphere paired for SPIM and control imaging purposes. Samples approximately 3 mm from the dorsal cortical surface were imaged by suspending them in agar in a syringe tube (SPIM) or mounting them on a microscope slide and sealing with agar (epifluorescence, confocal, and two-photon).

\begin{figure}[t]
	\centering
	\includegraphics[width=\columnwidth]{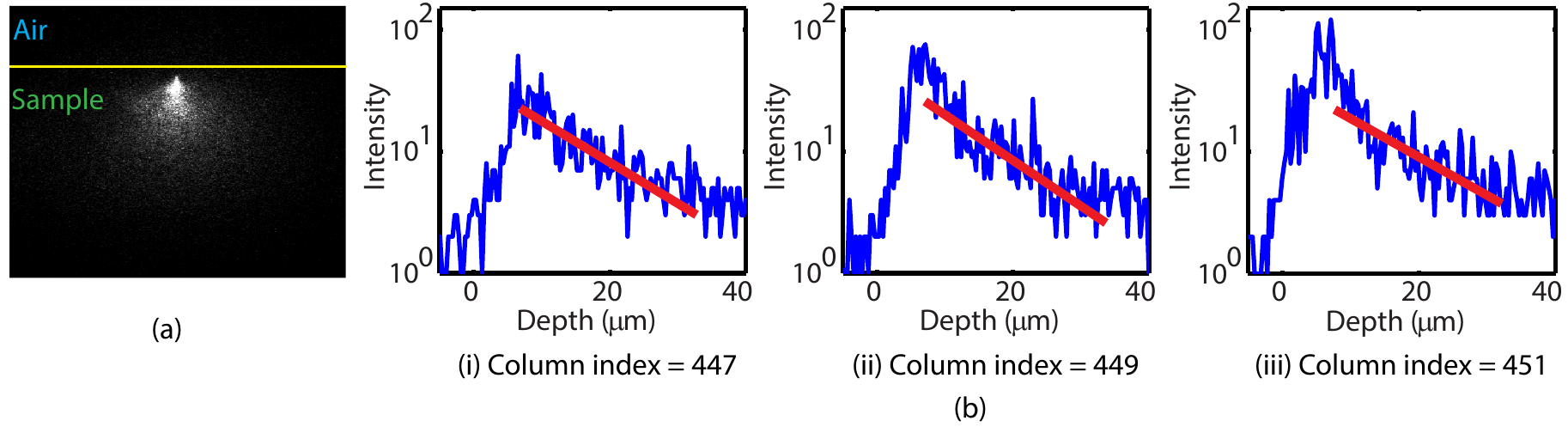}
	\caption{{\bf Experimental data to measure MFP.} (a) The laser beam is scattered as it goes through the medium (brain phantom). (b) For a length up to single scattering event, the laser intensity profile is going to decrease logarithmically with decay rate of 1/MFP. By measuring the rate of decay of the beam for various columns, we estimate the mean free path length of the sample.}
	\label{Fig:MFPMeasurement}
\end{figure}
\subsection{Measuring MFP for brain phantom}
We first compute MFP using Mie scattering theory. Later, we experimentally calculate the MFP and verify that the experimentally verified value is close to the theory. The mean free path length of the sample is given by \cite{mengual1999turbiscan}:
\begin{align}
	MFP = \frac{2d}{3\phi Q_s},
	\label{eq:MFPMieTheory}
\end{align}
where d is the mean diameter of the bead,   is the scattering efficiency factor (wavelength dependent), and   is the volume density of the scattering medium.

In the brain phantom sample we have used, we have   beads per mL and the average diameter of the bead is 1 $\mu$m. The scattering efficiency of polystyrene at 632.8 nm is 2.7784 \cite{gu2015microscopic}. Hence, using Eq.~\ref{eq:MFPMieTheory} , the MFP for the brain phantom is calculated to be 83.91 $\mu$m at 632.8 nm or 74.25 $\mu$m at 560 nm. The MFP for the half and quarter concentration brain phantoms used in our experiments is 148.50 $\mu$m and 297.00 $\mu$m.

However, as the preparation of phantom is not very reliable and is subject to procedural errors, such as wrong concentration of beads that may have perturbed the MFP of the sample, the actual MFP of the phantom can deviate from the calculations of the Mie theory. Therefore, we experimentally measured the MFP of each phantom sample employed in our experiments and used this experimentally measured MFP for normalizing the depth limits.

Fig.~\ref{Fig:MFPMeasurement}(a) shows the side view of laser scattering from the sample. The vertical profile of the image for various column pixels is shown in Fig.~\ref{Fig:MFPMeasurement}(b). Let $I(d)$  be intensity profile of a vertical plane, $\tau$ be the MFP, $d_s$ be the starting location of the beam. If we employ the zeroth order approximation of RTE, we get Beer-Lambertian law \cite{swinehart1962beer} and we have:
\begin{align}
	I(d) = \alpha  \exp \left(-\frac{d-d_0}{\tau} \right),
	\label{eq:MFPMieTheory}
\end{align}

\noindent where $d>d_0$ is the depth. $\alpha$ accounts for the intensity of the laser, quantum yield, and albedo. Taking logarithm on both sides, Eq.~\ref{eq:MFPMieTheory} becomes:
\begin{align}
	\log I(d) = \log \alpha -\frac{d}{\tau} + \frac{d_0}{\tau}.
	\label{eq:logMFPMieTheory}
\end{align}

Hence, the slope of $\log I(d)$  vs $d$ plot gives us $\displaystyle -\frac{1}{\tau}$ and is independent of both $\alpha$ and $d_0$. Therefore, we fit a straight line for the log measurements as shown in Fig.~\ref{Fig:MFPMeasurement}(b). 

There are multiple column pixels in Fig.~\ref{Fig:MFPMeasurement}. Naturally, the average of the MFPs of all the columns is considered a good estimate of MFP. However, the angle of the light ray will also play a role in the MFP estimate. Hence, as we deviate from the direct light ray, the MFP estimate will increase. However, for columns where the beam travels in a straight line, the MFP estimate is going to be more accurate. For the data in Fig.~\ref{Fig:MFPMeasurement}, this corresponds to columns 447 to 452. We took the mean value of MFP estimates of these columns as the MFP estimate of the sample. After measuring MFP of multiple samples of the same batch and averaging the results, the mean free path length of this brain phantom is found to be 75.03 $\mu$m . We also did same experiment for half concentration and quarter concentration brain phantoms, and the MFP is found to be 140.12 $\mu$m and 238.53 $\mu$m. We can observe that the experimentally measured MFP is very close to the Mie scattering theory. We use experimentally measured MFP as distance unit to compute depth limits in the paper. 

\subsection{Microscopic Imaging setups}
In this subsection, we will give step-by-step details of imaging a sample with the help of SPIM, epi-fluorescence, confocal, and two-photon microscopes that can help researchers in replicating the research detailed in this paper. 

\subsubsection{SPIM}
We built our SPIM setup based on the OpenSPIM platform \cite{pitrone2013openspim}. The laser power is maintained at 50 mW to ensure that the fluorophore beads are saturated. Saturation of fluorophores ensure that the depth limit is not a function of the input power. The distance between two slices of the fluorophore image is 6 $\mu$m. The camera frame is 1024 $\times$ 1344 pixels and the resolution is 0.67 $\mu$m/pixel. The samples are fixed within a syringe and hung on a 4D stage, which can move the sample in four dimensions very precisely (1.5 $\mu$m per step in translation and 1.8 degree per tick in rotation). As imaging deeper, the exposure time is increased to compensate the decrease of fluorophore signal due to scattering.
The thickness of light sheet determines the optical sectioning capability of SPIM. We imaged Ronchi ruling (100 lp/mm) illuminated by LED light to determine the lateral resolution of microscope, which is measured to be 0.6778 mm/pixel. To measure the light sheet thickness, we hung a small glass slide in the sample chamber of SPIM at an angle of 45 degrees. Hence, the light sheet is reflected directly into our camera system. The light sheet showed up as a thin line in the camera frame. The full-width-half-maxima (FWHM) of the cross section of the line is 11 pixels that translates to a physical distance of 7 $\mu$m, the thickness of our light sheet.
\subsubsection{Epi-fluorescence}
We used Nikon A1-Rsi confocal microscope system with its EPI mode as an epifluorescence microscope for measuring the depth limit. The sample was illuminated from the top with LED light filtered by blue filter as the fluorophore's excitation wavelength is 488 nm. The green emission light is detected at the top of sample by wide-field objective. By moving the sample along the detection axis, we image the sample slice by slice. Each slice corresponds to image of the portion of the sample that falls in the depth of field of objective. To compare with SPIM microscope, we maintained the distance between two slices of the fluorophore images as 6 $\mu$m  same as the distance between two slices of SPIM. As the focus volume goes deeper into the sample, the LED light power is increased to compensate the decrease of signal level due to scattering. The depth limit is the largest depth beyond which we are unable to identify fluorophores.
\subsubsection{Confocal microscopy}

We used the Nikon A1-Rsi confocal microscope for measuring the depth limit. The depth limit is dependent on the size of the pinhole \cite{Kempe:96}. A small pinhole gives the microscope the ability to reject the scattered light and hence increases the depth limit. However, below a pinhole size, the signal loss is high and will negatively influence the depth limit. The optimal pinhole size is found to be 1.3 optical units \cite{Kempe:96}. We set the pinhole size $\nu_p$ of our confocal microscope as 1.3 optical units to obtain the best possible depth limit. For the sake of comparisons, we also used a very large pinhole size ($\nu_p$ = 50  optical units) and computed the depth limits. We expressed the pinhole size in optical units as this is a common practice in the literature. We can compute the pinhole diameter $d_p$ given pinhole size ($\nu_p$), focal length ($f$), and radius of objective ($\alpha$), as $d_p = \nu_p \lambda f / \alpha \pi$.

Another important parameter that influences depth limits is the exposure duration. Ideally, the depth limits must be independent of the exposure duration. Hence, we maintain the laser intensity high enough to saturate the fluorophore and exposure duration long enough not to saturate the sensor. The equivalent of exposure duration for the confocal microscope is pixel dwell time. It is defined as the amount of time the focused laser beam rests on a single pixel illuminating it. It is usually a very small value whose magnitude is in the order of micro-seconds. At the shallowest region, we set the pixel dwell time as 2 $\mu$m and laser power as 1 mW to obtain a properly exposed image slice. Both pixel dwell time and laser power are increased as we image deeper to make up for weakening signal due to light scattering. The laser power is increased to 100 mW and the pixel dwell time is increased to 32 $\mu$s to ensure the fluorophore saturation while simultaneously ensuring that the signal received is strong enough.

To compare with SPIM microscope, we maintained the distance between two slices of the fluorophore images as 6 $\mu$m, same as the distance between two slices of SPIM. The depth limit is the largest depth beyond which we are unable to identify fluorophores. 

\subsubsection{Two-photon microscopy}
For two-photon microscope experiment, we used the ``Ultima In-Vivo 2P microscope" from Prairie Technology and  ``MaiTai HP DeepSee" pulse laser from Spectra-Physics. Two photons of infrared light are absorbed at focus point to excite fluorescent dyes. Using infrared light will minimize scattering in the tissue and achieve deep tissue penetration. We used the same sample for Epifluorescence, Confocal and Two-photon microscopes for comparison purposes. The excitation wavelength of fluorophore for single-photon imaging is 488 nm. Hence, the wavelength of infrared laser is set to 976 nm. By moving the sample along the detection axis, we image the sample slice by slice. We increased the laser power as we imaged deeper to compensate for the scattering effects. To compare with SPIM microscope, we also maintained the distance between two slices of the fluorophore images as 6 $\mu$m, same as the distance between two slices of SPIM. The depth limit is the largest depth beyond which we are unable to identify fluorophores.

\section{Discussions and Conclusions}
Using a combination of Monte Carlo simulations and experiments we have discovered that SPIM can image more than two times deeper than other single photon microscopy techniques like confocal and epifluorescence and nearly as deep as 2PM. The latter is somewhat surprising and points toward exciting applications of SPIM for imaging in scattering media.
This fact creates opportunities for compact inexpensive SPIM microscopes that could be mounted on the head of freely moving animals (similar to current miniature fluorescence microscopes \cite{ghosh2011miniaturized}). Additionally, these results motivate efforts to develop improved tabletop SPIM techniques for brain imaging like SCAPE \cite{bouchard2015swept} where fast image acquisition rates are complemented by imaging depths that exceed other single photon techniques. 

Our results also inform the design of future SPIM techniques for imaging in scattering media. Namely we found that a combination of low noise sensors and thin light sheets can dramatically increase the imaging depth to approximately 15 MFPs according to our simulations. This result motivates future work to design systems capable of producing thin light sheets in scattering media perhaps using adaptive optics \cite{bourgenot20123d} or integrated photonics that can control the wavefront of the illumination source.

Overall we believe our results highlight SPIM as an exciting technology for imaging in scattering media. In addition to its well-known merits for high-speed image acquisition and low photobleaching rates, SPIM can image exceptionally deep compared to previously demonstrated single photon microscopy techniques.

\section*{Acknowledgments}
The authors would thank Chris Metzler, Jason Holloway, and Austin Wilson for proof-reading the document. The authors acknowledge the support from NSF 1502875, 1527501, 1308014, ONR N00014-15-1-2735, N00014-15-1-2878, AFOSR FA9550-12-1-0261.

\end{document}